\documentclass[letterpaper,twocolumn,10pt]{article}
\usepackage{authblk}
\usepackage{usenix,epsfig,endnotes}
\usepackage{booktabs}
\usepackage{fontawesome}
\usepackage{array}
\usepackage{multirow}
\usepackage{svg}
\usepackage{subcaption}
\usepackage{tikz}
\usepackage{amsmath}
\usepackage{listings}
\usepackage{filecontents}
\usepackage{xspace}
\usepackage{tabularray}
\usepackage[colorinlistoftodos,prependcaption,textsize=tiny]{todonotes}
\usepackage{enumitem}

\setlist[enumerate]{leftmargin=5mm}
\definecolor{codegreen}{rgb}{0,0.6,0}
\definecolor{codegray}{rgb}{0.5,0.5,0.5}
\definecolor{codepurple}{rgb}{0.58,0,0.82}
\definecolor{backcolour}{rgb}{0.95,0.95,0.92}
\definecolor{pinkcolour}{HTML}{de90c0}

\lstdefinestyle{mystyle}{
    backgroundcolor=\color{backcolour},   
    commentstyle=\color{codegreen},
    keywordstyle=\color{magenta},
    numberstyle=\tiny\color{codegray},
    stringstyle=\color{codepurple},
    basicstyle=\ttfamily\footnotesize,
    breakatwhitespace=false,         
    breaklines=true,                 
    captionpos=b,                    
    keepspaces=true,                 
    numbers=left,                    
    numbersep=5pt,                  
    showspaces=false,                
    showstringspaces=false,
    showtabs=false,                  
    tabsize=2
}

\newcounter{reqcntr}

\newcounter{reqcntr_sub}[subsection]

\definecolor{alizarin}{rgb}{0.82, 0.1, 0.26}

\definecolor{redish}{RGB}{234,107,102}
\newcommand*\negcircnum[1]{\tikz[baseline=(char.base)]{%
            \node[white,shape=circle,fill=redish,draw,inner sep=1pt] (char) {\color{white}\sffamily #1};}}

\newcommand*\negcircnumreq[1]{\tikz[baseline=(char.base)]{%
            \node[white,shape=circle,fill=blue,draw,inner sep=1pt] (char) {\color{white}\sffamily #1};}}

\newcommand*\negcircnumstep[1]{\tikz[baseline=(char.base)]{%
            \node[white,shape=circle,fill=pinkcolour,draw,inner sep=1pt] (char) {\color{white}\sffamily #1};}}

\newcommand*{\requirement}[1]{
    \vspace{0.3em}
    \stepcounter{reqcntr}
    \stepcounter{reqcntr_sub}
    \noindent\textbf{\negcircnumreq{R\arabic{reqcntr}}} \textit{#1}.
    }

\newcommand*{\subpar}[2]{
    \vspace{#1em}
    \noindent
    \textbf{#2}
}

\author[1]{Mikhail Khalilov}
\author[1]{Marcin Chrapek}
\author[1]{Siyuan Shen}
\author[1]{Alessandro Vezzu}
\author[2]{Thomas Benz}
\author[1]{\authorcr Salvatore Di Girolamo}
\author[1]{Timo Schneider}
\author[1,3]{Daniele De Sensi}
\author[2]{\authorcr Luca Benini}
\author[1]{Torsten Hoefler}
\affil[1]{SPCL, D-INFK, ETH Zurich}
\affil[2]{IIS, D-ITET, ETH Zurich}
\affil[3]{Department of Computer Science, Sapienza University of Rome}
{
    \makeatletter
    \renewcommand\AB@affilsepx{: \protect\Affilfont}
    \makeatother

    \affil[ ]{}

    \makeatletter
    \renewcommand\AB@affilsepx{, \protect\Affilfont}
    \makeatother
}

\begin{document}

\date{}

\title{\Large \bf OSMOSIS: Enabling Multi-Tenancy in Datacenter SmartNICs}

\maketitle

\thispagestyle{empty}

\subsection*{Abstract}
Multi-tenancy is essential for unleashing SmartNIC's potential in datacenters. Our systematic analysis in this work shows that existing on-path SmartNICs have resource multiplexing limitations. For example, existing solutions lack multi-tenancy capabilities such as performance isolation and QoS provisioning for compute and IO resources. Compared to standard NIC data paths with a well-defined set of offloaded functions, unpredictable execution times of SmartNIC kernels make conventional approaches for multi-tenancy and QoS insufficient. We fill this gap with OSMOSIS, a SmartNICs resource manager co-design. OSMOSIS extends existing OS mechanisms to enable dynamic hardware resource multiplexing of the on-path packet processing data plane. We integrate OSMOSIS within an open-source RISC-V-based 400Gbit/s SmartNIC. Our performance results demonstrate that OSMOSIS fully supports multi-tenancy and enables broader adoption of SmartNICs in datacenters with low overhead.

\begin{figure}[ht!]
    \centering
    \includegraphics[width=\columnwidth]{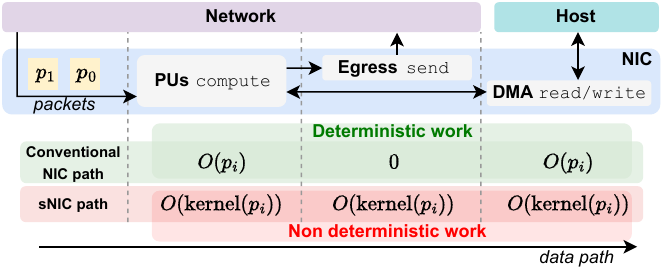}
    \caption{A predictable NIC data path versus the unpredictable sNIC kernel execution.}
    \label{fig:overview}
\end{figure}
\section{Introduction}
Network data plane design has undergone two decades of exciting research, leading to the achievement of sub-microsecond packet processing host latency \cite{rdmaguidelines, hoiland2018express, belay2014ix, peter2015arrakis, prekas2017zygos, kalia2018datacenter, kaffes2019shinjuku, fried2020caladan, farshin2021packetmill, ibanez2021nanopu, miano2022domain, seyedroudbari2023turbo}. SmartNICs (sNICs) have further improved processing times by enabling direct in-network packet processing, thereby reducing data movement \cite{ivanov2021data}. sNICs started a trend in datacenter networking acceleration \cite{vahdat2021next, katsikas2021you} similar to the GPU trend in high-performance computing~\cite{wang2022fpganic}.

sNICs enable running \textit{kernels} on programmable, energy-efficient cores tailored for packet processing and integrated within the host network interface card (NIC) System-on-Chip (SoC). These cores are attached directly (i.e., \textit{on-path}) to the datacenter Ethernet or InfiniBand link~\cite{attig2011400, le2017uno}. Such a design reduces the latency of some applications since the sNIC can process the packets in the network~\cite{liu2019e3} and reply directly without moving the packets to/from the host OS networking stack~\cite{haecki2022diagnose, agarwal2022understanding}. This enables the offload and acceleration of several workloads such as distributed learning gradients aggregation~\cite{wang2022fpganic, swamy2022taurus}, disaggregation and storage~\cite{minturn2015nvm, gao2016network, min2021gimbal, kim2021linefs, guo2022clio}, Key-Value Stores (KVS)~\cite{sun2022skv, pourhabibi2021cerebros, yuan2023rambda}, Remote Procedure Calls (RPCs)~\cite{liu2019offloading, lazarev2021dagger, yan2020p4,borromeo2022fpga, rivitti2023ehdl}, network protocols and telemetry~\cite{yang2020making, mogul2003tcp, yan2020p4, borromeo2022fpga, eran2022flexdriver, ibanez2022enabling, snictrafficanalysis, chang2023learned}.

Network resources in a datacenter are multiplexed between tenants through a virtualization layer~\cite{dean2013tail,mudigonda2011netlord, blocher2021switches, justitia, kumar2019picnic}. However, processing user code by sNICs brings a set of considerable resource management issues. As Figure \ref{fig:overview} shows, NICs have three resources that must be multiplexed: compute, Direct Memory Access (DMA) bandwidth, and egress bandwidth. The traditional NIC data path only forwards packets to host memory and executes simple operations with a \textit{predictable} and \textit{bounded} complexity. Typically, the number of incoming bytes equals the number of outgoing bytes, and NICs do not run any elaborate processing on them. In contrast, sNICs can execute \textit{unpredictably} complex stateful offloads~\cite{pismenny2021autonomous}. For example, heavily used in machine learning~\cite{ben2019demystifying} Allreduce operates on the payload and is compute-bound, while storage offloading predominantly accesses host memory and is DMA/IO bound. sNICs need to operate on \textit{uncoordinated}, \textit{non-deterministic}, and \textit{concurrent} data streams while meeting Service Level Objective (SLO) policies set by the administrator.

Achieving a fair resource multiplexing for sNICs is challenging. sNICs combine characteristics of an accelerator, such as a GPU, and a traditional NIC. While this provides the aforementioned benefits, the resource management of neither is directly applicable due to the unique sNIC requirements (Section~\ref{sec:requirements}). Conventional RDMA NICs (rNICs) have bounded and predictable workloads (e.g., atomics, scatter-gather RDMA reads/writes) and often use link bandwidth allocation as a \textit{"just enough"} mechanism for resource isolation and Quality-of-Service (QoS) measure between tenants. Although rNICs exhibit bounded and foreseeable behavior, achieving fairness is challenging~\cite{weibaiunderstanding} even within their simpler than sNIC context.
In contrast, accelerators fall entirely under the governance of the host OS, which oversees all active kernels~\cite{korolija2020abstractions, khawaja2018sharing}. These accelerators neither generate nor receive events beyond instructions from accelerated applications, setting them apart from sNICs capable of executing arbitrary kernels independently of the host's involvement.

Furthermore, for sNICs to sustain the sub-\textit{nano}second packet arrival intervals at fully utilized 400Gbit/s link (Section~\ref{sec:requirements}, \cite{gao2022gearbox}), resource multiplexing must be conducted fast. On-path sNICs have much stricter compute and buffering constraints than traditional NICs and accelerators due to the packet rate and the three multiplexed resources (compute, DMA, and egress). This issue is even more critical as network rates constantly increase and are expected to exceed Terabit per second by 2025~\cite{cai2022towards, hoefler2023data, ethernetroadmap, ultraethernet}.

A common approach to effectively manage processing at high packet arrival rates involves implementing resource management in hardware~\cite{anderson1993high, agrawal2020intel, gao2022gearbox}. This is usually accomplished through scheduling policies such as Weighted Round Robin (WRR), which divide link bandwidth among tenants~\cite{weibaiunderstanding, dong2012high, dong2008sr}. However, because sNICs have varying application kernel requirements, incorporating WRR for compute resource allocation can lead to unfairness. For example, as we show in Section~\ref{sec:requirements}, if one application (e.g., Allreduce) is compute-bound and takes twice as much compute time as a non-compute-bound application (e.g., KVS), the former will be able to process twice as many bytes. Other recently proposed methods for compute isolation in sNICs are not optimal for all scenarios as they are either non-work conserving~\cite{grant2020smartnic} or rely on the host CPU as a fallback path~\cite{liu2019offloading}.

We tackle these issues by introducing OSMOSIS (\underline{O}perating \underline{S}yste\underline{m} Supp\underline{o}rt for \underline{S}treaming \underline{I}n-Network Proce\underline{s}sing) (Section~\ref{sec:osmosis}). OSMOSIS is a lightweight sNIC management layer that supports performance-critical data-plane management in hardware and non-critical management tasks in a flexible software runtime. OSMOSIS is a fair, work-conserving sNIC resource manager that requires minimal hardware footprint and employs expressive yet simple Service Level Objective (SLO) semantics. In OSMOSIS, the sNIC is exposed to the tenant as Single-Root Input/Output Virtualization (SR-IOV) Virtual Function (VF). This allows the administrator to allocate proportionally more \textit{compute processing units}, \textit{egress bandwidth}, and \textit{DMA bandwidth} to VFs associated with high-priority tenants.

We implement (Section~\ref{sec:implementation}) and evaluate (Section~\ref{sec:evaluation}) OSMOSIS on top of one of the available open-source on-path sNIC architectures, PsPIN \cite{hoefler2017spin, di2021risc}. PsPIN is based on energy-efficient silicon-proven RISC-V cores. In our setup, PsPIN is the hardware backbone for packet processing using kernels written in C. Our performance evaluation focuses on typical datacenter workloads such as storage IO and in-network Allreduce, and shows that OSMOSIS provides comprehensive support for multi-tenancy without sacrificing performance.

In summary, we make the following contributions:
\begin{enumerate}
\item \textit{sNIC multi-tenancy:} We show typical multi-tenancy sNIC problems and define a set of requirements for high-performance sNICs. These requirements serve as a guideline for developing sNICs that can meet the needs of diverse workloads and tenant environments (Section \ref{sec:requirements}).

\item \textit{OSMOSIS:} We introduce OSMOSIS, a lightweight open-source sNIC resource manager based on fair and work-conserving scheduling policies. OSMOSIS is a minimal hardware footprint solution to the problem of fair and efficient resource sharing in multi-tenant sNICs with diverse application needs (Section \ref{sec:osmosis}).

\item \textit{Evaluation:} We implement OSMOSIS in an open-source on-path 400Gbit/s sNIC by extending it with schedulers and a control path prototype (Section \ref{sec:implementation}). We use this implementation to verify and evaluate OSMOSIS. We demonstrate how it solves the defined sNIC problems and handles multi-tenant applications fairly with varying resource requirements while minimizing tail latency (Section \ref{sec:evaluation}).

\end{enumerate}

\section{Background}\label{sec:background}

From the system's perspective, we abstract out the sNIC as a packet processing accelerator between the network fabric and the host CPU, GPU, or FPGA. Existing sNICs can be classified broadly into two categories: \textit{off-path} and \textit{on-path}~\cite{liu2019offloading}. 

Off-path sNICs add an entire CPU complex to the network card, often running a full operating system (e.g., Linux). This design enables a management plane based on receive side scaling (RSS) to be conveniently implemented~\cite{rss, belay2014ix, prekas2017zygos}. However, they often suffer from lower performance in terms of latency, bandwidth, and packet processing rates due to their system design, which closely resembles the CPU-centered host architecture (e.g., Broadcom Stingray and Nvidia Bluefield data processing units (DPUs) both feature ARM SoCs with PCIe and DRAM).

On-path sNICs share packet input buffers with \textit{processing units} (PUs) tailored for highly-parallel packet processing (e.g., LiquidIO~\cite{liquidio}, Netronome~\cite{netronome}, PsPIN~\cite{di2021risc}, Data Path Accelerator (DPA) introduced in Bluefield 3 DPU~\cite{bf3-flexio, bf3}). On-path sNICs typically provide programming API for writing \textit{kernels} that process traffic on PUs, on per-packet (PsPIN~\cite{di2021risc}) and/or per-message granularity (Bluefield-3 FlexIO API~\cite{bf3-flexio}). PUs typically feature three layers of the memory hierarchy, e.g., L1 single-cycle access scratchpad, L2 memory with access latency of 15-50 cycles, and host side memory (either off-path SoC or host CPU memory). L1 and L2 memories could be organized as multi-level caches (e.g., LiquidIO) or be explicitly managed by the user (e.g., PsPIN).

To our knowledge, OSMOSIS is the first solution to achieve fair resource multiplexing for on-path sNICs in a multi-tenant context. We selected one of the possible synthesizable open-source on-path sNIC implementations available in the literature, namely, PsPIN. PsPIN is open-source, based on energy-efficient silicon-proven RISC-V cores, and allows users to write packet processing kernels in C and explicitly manage sNIC memory~\cite{di2021risc}. OSMOSIS could have been equivalently implemented in any other on-path framework~\cite{bf3-flexio, netronome,liquidio}. For example, we discuss how OSMOSIS can be supported with BlueField-3 DPA in Section~\ref{sec:bf3dpa} 

\subsection{Challenges of Resource Isolation}\label{sec:lack_of_isolation}

\begin{figure}[t!]
    \centering
    \includegraphics[width=\columnwidth]{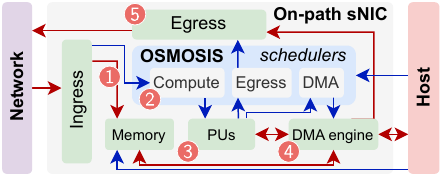}
    \caption{Schematic overview of on-path sNIC architectures. Red arrows indicate the data path and blue arrows correspond to the control/management path.}
    \label{fig:template_arch}
\end{figure}

We generalize on-path sNIC architecture in Figure~\ref{fig:template_arch}. Packets decoded from the sNIC physical layer (e.g., Ethernet MAC) arrive at the sNIC inbound engine~\negcircnum{1} and are initially stored at the L2 packet buffer organized as a set of per-application first-in-first-out (FIFO) queues. Next~\negcircnum{2}, packets are scheduled for processing on available PUs where kernel execution is initiated~\negcircnum{3}. Kernels execute using three resources: PUs, DMA, and Egress bandwidth. Each application uses these resources differently (e.g., compute- or IO-bound) depending on its needs. In general, these resources can be used as follows:

\begin{itemize}
\itemsep0em
\item[\negcircnum{3}] PUs: computing (e.g., hashing the packet header or summing values in an Allreduce reduction);
\item[\negcircnum{4}] DMA engine: transferring data to read/write in sNIC memory (e.g., KVS cache in sNIC L2 memory) or host memory (e.g., KVS cold storage);
\item[\negcircnum{5}] Egress engine: sending packet replies (e.g., reply to a read request with a value from the KVS cache). 
\end{itemize}

Metrics to measure the quality of resource multiplexing by datacenter tenants, known as Service-Level Objectives (SLOs), are typically tied to the conventional NIC path displayed in Figure~\ref{fig:template_arch} by considering tail latency \cite{dean2013tail} and throughput \cite{singla2014high, namyar2021throughput}. However, these SLOs do not consider the sNIC data path with its unique resource multiplexing discussed in Section~\ref{sec:requirements}, such PU time, tail latency of DMA over host interconnect, and buffer space. Existing proposals have only partially addressed this issue by introducing performance isolation mechanisms, such as multi-level packet scheduling~\cite{gao2022gearbox, liu2019offloading, stephens2019loom} and static resource allocation~\cite{grant2020smartnic} of shared resources (see Section~\ref{sec:comparison}). Yet, due to the kernels' dynamic and unpredictable nature, static assignments do not solve the problem. OSMOSIS fills this gap by \emph{providing bounded guarantees for the sNIC resource availability to tenants using dynamic resource multiplexing}. 

\section{Multi-Tenant sNICs}\label{sec:requirements}
Datacenter applications differ in their resource requirements, thus, leading to different resource multiplexing bottlenecks. Our quantitative analysis highlights these issues in multi-tenant setups of existing sNIC stacks~\cite{di2021risc, bf3-flexio}, yielding sNIC multi-tenancy requirements. These insights directly led to the microarchitectural and software choices for OSMOSIS. We use a 400 Gbit/s link for all experiments (more details on the setup in Section~\ref{sec:evaluation}).

\subpar{1}{Per-packet time budget (PPB):} While studies of datacenter traffic show that only a fraction of the established connections actively exchange data at any given time~\cite{clusterdata:dcburst, clusterdata:fbdataset, clusterdata:private}, they can still saturate the link bandwidth. To analyze the implications of this for sNICs we define per-packet time budget (PPB) using PU count $N$, packet size $P$, and link bandwidth $B$ as $PPB(N, P, B) = N\times(P / B)$. In this case, we model the sNIC as a $M/M/m$ queue where PPB defines the condition which needs to be satisfied for the queue to be stable~\cite{bolch1998queueingnetworks}\footnote{$1 / \lambda = P / B$, $m = N$, to achieve $\rho < 1$, $1/ \mu > N \cdot P / B$, where PPB = $1 / \mu$.}. To be more specific, PPB represents how long the sNIC can process a packet until the next one arrives, assuming a fully utilized link. If PPB is exceeded, the per-application ingress queue will eventually fill up during transient traffic bursts leading to packet drops or falling back to link flow control (e.g., PFC~\cite{zhu2015dcqcn}) and a possible violation of per-VF SLO policy.

\begin{figure}[]
    \centering
    \includegraphics[width=\columnwidth]{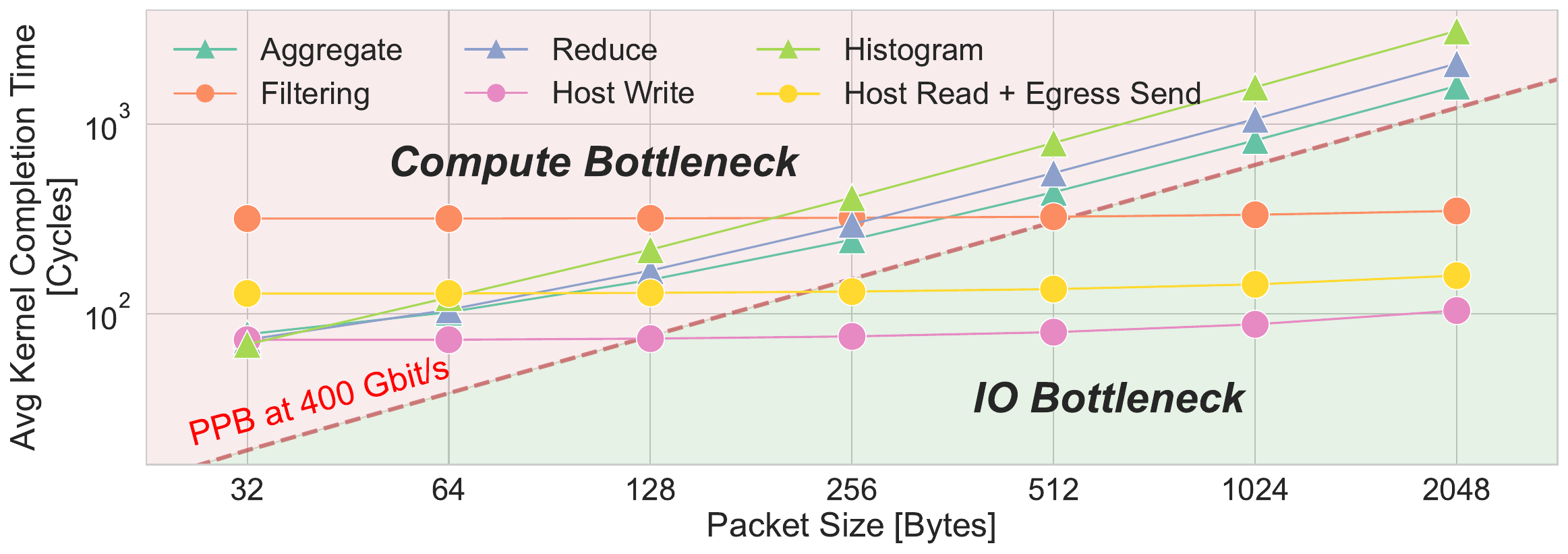}
    \caption{sNIC core (PU) processing time needed to serve $1$ packet for common sNIC kernels. Workloads with triangle markers are compute-bound, and circular markers are IO-bound. All workloads with $\leq64$B packet size (including 28 bytes IPv4/UDP-header) exceed PPB showing congestion at PUs when link bandwidth is fully utilized. Note that our setup supports Ethernet payload sizes below $64$B to accommodate custom interconnects~\cite{ibspec}.}
    \label{fig:ppb}
\end{figure}

Figure~\ref{fig:ppb} compares service times of IO-- and compute-bound workloads with theoretical PPB assuming that tenant workloads fit one packet and that the sNIC has only one tenant. We observe that all workloads with packet size $\leq64$~Bytes fail to fit in PPB. Compute-bound workloads (i.e., Aggregate, Reduce, Histogram) whose execution time scales linearly with packet payload length exceed the PPB for all packet sizes bottlenecking the PUs. Notably, IO-bound kernels above $256$~Bytes (i.e., DMA writes/reads, Egress packet sends) fit PPB as they avoid PU congestion but are bottlenecked by the link bandwidth. However, as we will demonstrate, \textit{IO-bound workloads are sensitive to DMA transfer contention on the host interconnect}.

\begin{figure}[]
    \centering
    \includegraphics[width=\columnwidth]{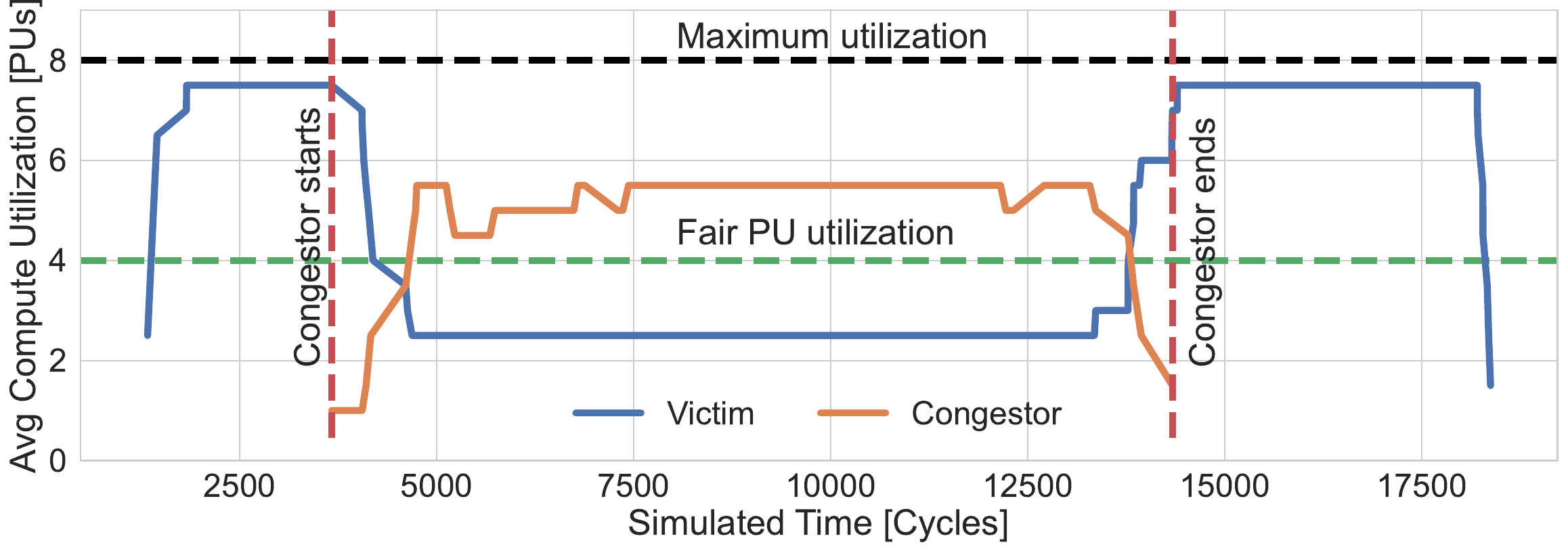}
    \caption{\textit{Congestor} and \textit{Victim} tenants' flows with equal priorities are mapped to two different SR-IOV VFs with equal shares of Ingress bandwidth. With the round-robin scheduling of per-flow queues, the \textit{Congestor} tenant with 2$\times$ higher compute cost per packet occupies a proportionally larger number of cores than the \textit{Victim} tenant.}
    \label{fig:hpu_unfair}
\end{figure}

\subpar{1}{PU contention:} While a single tenant can cause pressure on the ingress queue and contention of PUs, multiple tenants can lead to unfairness. For example, consider two compute-bound tenants with different requirements. One of them, the \textit{Congestor}, has twice as large compute cost per packet as the other, the \textit{Victim}, leading to twice as many cycles on PU to finish the kernel. During the burst, \textit{Congestor} and \textit{Victim} push packets at the corresponding per-application (per-VF) queues at the same ingress rate. As Figure~\ref{fig:hpu_unfair} shows, using the conventional round robin (RR) scheduling of per-application queues across $8$ sNIC PUs, the \textit{Congestor} uses $2\times$ the PUs used by the \textit{Victim}.

\requirement{sNIC manager should fairly allocate compute components (e.g., PUs, cryptographic accelerators) while serving tenants with different compute costs per packet}

\subpar{1}{Egress and DMA engines contention:} Similarly, as the compute-bound kernels cause contention on PUs, IO-bound kernels can lead to contention on the appropriate DMA or egress engines. IO-bound kernels running on different PUs can simultaneously initiate IO requests through the same sNIC engines, e.g., DMA requests from a KVS application. In case the underlying interconnect (e.g., PCIe or AXI~\cite{restuccia2019your}) is blocking and lacks the support of QoS provisioning, \textit{the issue of multiple concurrent requests may result in Head-of-Line (HoL) blocking}~\cite{agarwal2022understanding}. 

\begin{figure}[]
    \centering
    \includegraphics[width=\columnwidth]{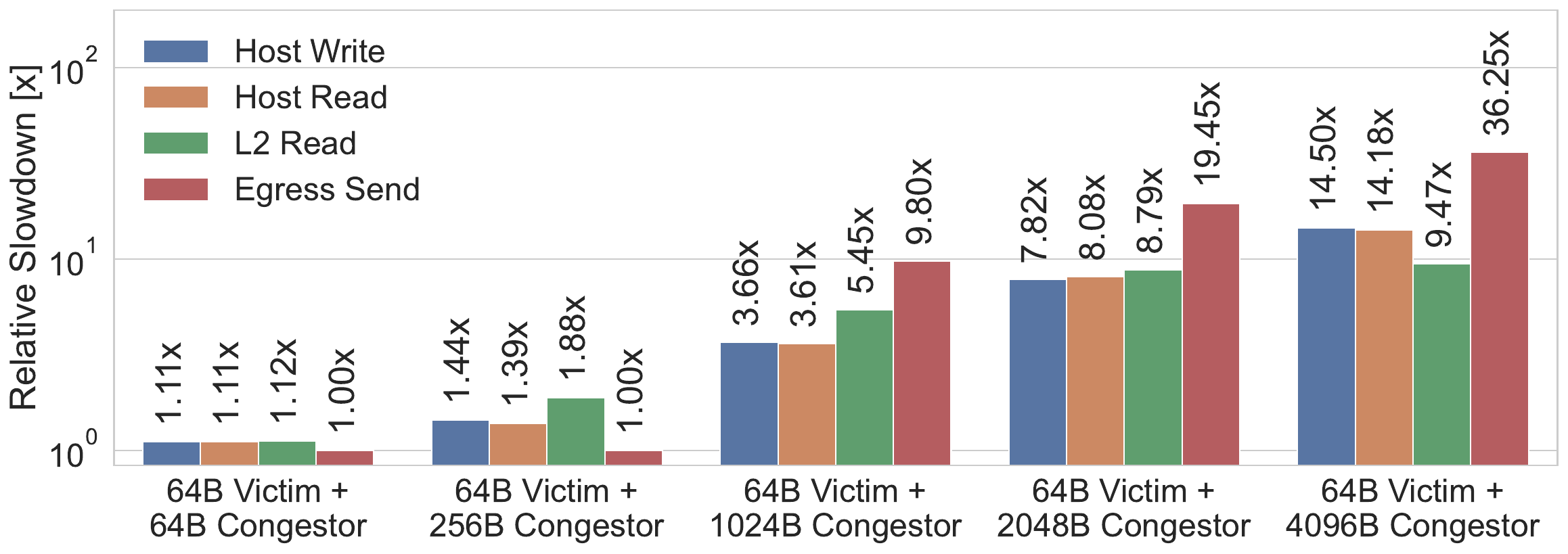}
    \caption{Slow-down of various IO operations (e.g., DMA and sending packets to Egress) initiated by the tenant's kernel results in HoL-blocking small requests due to underlying IO path contention.}
    \label{fig:dma_unfair}
\end{figure}

For example, consider two IO-bound tenants with different IO requirements. The \textit{Victim} has constant 64B packets, while the \textit{Congestor} increases its packet size from 64B to 4096B. As Figure~\ref{fig:dma_unfair} shows, the contention on the IO engine leads to an order of magnitude higher latency of the \textit{Victim}'s messages without considerably affecting the \textit{Congestor}'s flow. This unfairly increases the latency of one of the tenants by 4-15$\times$.

\requirement{sNIC manager should fairly allocate DMA and egress bandwidth (e.g., using AXI and PCIe) between running kernels and be resilient to HoL-blocking}

\subpar{1}{Memory management:} Applications have diverse memory runtime needs, with dynamic memory allocation causing an unknown \textit{a priori} memory consumption. In extreme cases, a tenant could monopolize all sNIC memory, e.g., L1 packet buffers, resulting in HoL-blocking for others. Introducing virtual memory (paging) semantics could lead to substantial memory access overheads, as each page fault significantly amplifies memory access latency~\cite{273711}.

\requirement{sNIC manager should fairly allocate memory using lightweight allocation strategies defined in the control plane}

\subpar{1}{Scheduling overhead:} Existing \textit{software} packet processing data paths~\cite{belay2014ix,prekas2017zygos,farshin2021packetmill} were designed for off-path sNICs or conventional host processing. As recent studies show~\cite{kaffes2019shinjuku} \textit{effectiveness} of kernel execution scheduling in terms of achieved maximum utilization while running on off-path sNICs supported by OS's like Linux is driven by the latency of context switching\cite{kaffes2019shinjuku, fried2020caladan}. PU cycles are wasted during context switching to transition between the kernel states. We benchmark context-switching of Linux running on host and off-path sNIC (Bluefield-2 ARM SoC). We compare these to the state-of-the-art Caladan scheduler we ported to the ARM ISA~\cite{fried2020caladan}. For reference, we also show the context switching latency of PULP cores as implemented in PsPIN used to evaluate OSMOSIS. Notably, we observe that the context switching latencies we report in Table~\ref{tab:ctx_lat} are higher or of the same order of magnitude as the PPB from the analysis presented in Figure~\ref{fig:ppb}.

\requirement{Data path performance should not be impacted by overheads stemming from software scheduling policies, providing low-latency scheduling of kernel execution}

\begin{table}[]
\resizebox{\columnwidth}{!}{%
\begin{tabular}{@{}cccccc@{}}
\toprule
\textbf{PU}                                                                   & \textbf{Frequency} & \textbf{ISA}    & \textbf{Linux} & \textbf{Caladan} & \textbf{RTOS} \\ \cmidrule(r){1-3} \cmidrule{4-6}
Host Ryzen 7 5700                                                    & 3.8GHz    & x86    & 28576 & 211     & --   \\ \addlinespace[0.5em]
BF-2 DPU A72                                                         & 2.5GHz    & ARMv8  & 13250 & 192     & --   \\ \addlinespace[0.5em]
\begin{tabular}[c]{@{}c@{}}PULP cores~\cite{pulp2021latency}\\ (used in PsPIN)\end{tabular} & 1GHz      & RISC-V & --    & --      & 121  \\ \bottomrule
\end{tabular}%
}
\caption{Average latency of context switching between 2 processes. Measurements shown in PU cycles scaled to 1 GHz (i.e., $1$ ns/cycle).}
\label{tab:ctx_lat}
\end{table}

\subpar{1}{Control path priority:} If a tenant on the sNIC exceeds compute or time budgets, an immediate response is needed from the host's control plane for \textit{control traffic}, e.g., it needs to be handled within the error path of the application running on the host CPU or off-path sNIC cores. However, communication between sNIC and host uses system interconnect (e.g., PCIe), typically adding an overhead of $0.5$ -- $3$ usec per read/write request. Congestion in the interconnect (Figure~\ref{fig:dma_unfair}, ~\cite{agarwal2022understanding}) can lead to HoL-blocking of control traffic and unpredictable packet processing.

\requirement{sNIC accelerated packet processing should prioritize control-path traffic}

\subpar{1}{QoS API:}
NIC capabilities are exposed to tenants through a virtualization layer (OS hypervisor) that provides an illusion of full resource ownership. SR-IOV is a standardized extension for the PCIe interconnect and a conventional way to implement NIC virtualization. It is utilized in many conventional industry-standard NICs, e.g., ConnectX and BlueField NICs. In SR-IOV, each NIC physical function (PF) (such as TX and RX capabilities) is multiplexed between several virtual functions (VFs). Each VF is exposed to the tenant through an OS hypervisor as a stand-alone PCIe NIC. To our knowledge, existing production rNICs and sNICs support only Ingress and Egress bandwidth allocation on the basis of VFs and not compute or DMA resources.

\requirement{sNIC management plane should support conventional QoS provisioning mechanisms for all types of resources}

\begin{figure}[th!]
    \centering
    \includegraphics[width=\columnwidth]{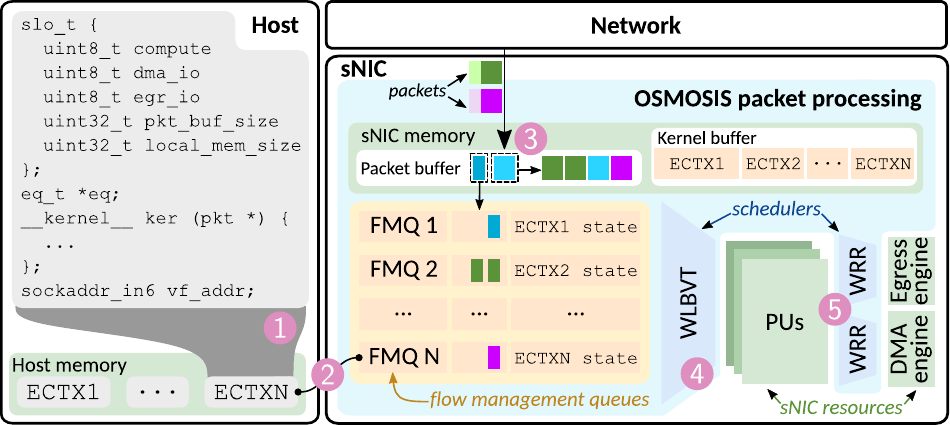}
    \caption{Abstract model of OSMOSIS-enabled sNIC. Packets are mapped by Matching Engine to FMQs and dispatched for execution by the scheduler.}
    \label{fig:snic_stack}
\end{figure}

\section{OSMOSIS}\label{sec:osmosis}

We present OSMOSIS in Figure~\ref{fig:snic_stack}. We begin with a high-level overview of how OSMOSIS manages the three competing sNIC resources and satisfies the multi-tenancy requirements outlined in the previous section. We then demonstrate how this is achieved by dividing the system into two components. The first is a non-critical, flexible software control plane that handles management tasks and runs on the host CPU or off-path sNIC cores. The second is a performance-critical data plane scheduler designed specifically to support SLO policy enforcement and integrate within the on-path sNIC SoC. Within this section, each part is explained in depth.

\begin{table}[th!]
\resizebox{\columnwidth}{!}{%
\begin{tabular}{@{}rcccc@{}}
\toprule
\multicolumn{1}{l}{}                                             & \textbf{PUs}                                                                               & \textbf{DMA}                                                                                                          & \textbf{Egress}                                                                            & \textbf{Memory}                                                                            \\  \cmidrule(r){2-2} \cmidrule(r){3-3} \cmidrule(r){4-4} \cmidrule{5-5} \addlinespace[0.5em]
Scheduler                                                        & WLBVT                                                                             & WRR                                                                                                          & WRR                                                                               & Static                                                                            \\ \addlinespace[0.5em]
SLO knob                                                         & \begin{tabular}[c]{@{}c@{}}Priority\\ Kernel cycle limit\end{tabular}             & Priority                                                                                                     & Priority                                                                          & Allocation size                                                                   \\ \addlinespace[0.5em]
\begin{tabular}[c]{@{}r@{}}Fulfilled\\ requirements\end{tabular} & \textbf{\negcircnumreq{R1}}\textbf{\negcircnumreq{R4}}\textbf{\negcircnumreq{R6}} & \textbf{\negcircnumreq{R2}}\textbf{\negcircnumreq{R4}}\textbf{\negcircnumreq{R5}}\textbf{\negcircnumreq{R6}} & \textbf{\negcircnumreq{R2}}\textbf{\negcircnumreq{R4}}\textbf{\negcircnumreq{R6}} & \textbf{\negcircnumreq{R3}}\textbf{\negcircnumreq{R4}}\textbf{\negcircnumreq{R6}} \\ \bottomrule
\end{tabular}%
}
\caption{OSMOSIS resource management principles with all six fulfilled multi-tenancy requirements.}
\label{tab:osmosis_principles}
\end{table}

\subsection{High-level Overview}

\subpar{0.5}{\negcircnumstep{1} Flow execution context creation:} To utilize sNIC packet processing, tenants create a flow \textit{execution context} (ECTX). ECTX encapsulates the flow processing state, such as the SLO policy and the packet processing \textit{kernel}, a piece of code compiled for the target PU architecture and describing the actions for each packet destined for the flow. 

\subpar{0.5}{\negcircnumstep{2} ECTX initialization:} After the tenant provides the basic elements of an ECTX, OSMOSIS instantiates it. It allocates a virtualized sNIC interface through the host OS hypervisor and associates it with a tenant IP address and SLO policy. It also sets up the IOMMU to allow kernel access to specific host pages, \textit{statically} allocates on sNIC memory and loads the kernel binary into sNIC memory.

\subpar{0.5}{\negcircnumstep{3} Matching packets to flow management queue:} The sNIC matching engine filters packets that require sNIC processing. All incoming packets are matched against the three-tuple (in case of UDP) or five-tuple (in case of TCP) of active sNIC ECTXs. Once matched, \textit{packet descriptors} (e.g., pointer to packets in sNIC memory) are stored at one of the \textit{flow management queues} (FMQs). FMQs store all information regarding an active flow ECTX on the sNIC hardware. FMQs are organized as FIFO queues of packet descriptors with an additional memory state to store running execution information (e.g., BVT metric).

\subpar{0.5}{\negcircnumstep{4} PU scheduling:} Once a PU becomes available, OSMOSIS schedules the packet at the head of one of the FMQs. To achieve fair PU allocation, OSMOSIS implements a centralized, non-preemptive scheduler inspired by the Borrowed Virtual Time (BVT) policy~\cite{duda1999borrowed, kaffes2019shinjuku}. BVT aims to allow each tenant to obtain the same amount of access time to the scheduled resource by keeping track of their past usage. OSMOSIS FMQ scheduler \textit{allocates sNIC PUs to FMQs with the smallest priority-adjusted past PU usage measured in cycles} while maintaining the SLO policy specified by the sNIC administrator, such as the upper per-FMQ PU cycle limit.

\subpar{0.5}{\negcircnumstep{5} Kernel execution and IO management:} Upon loading the packet into local PU memory, the PU can process it using the relevant kernel. As seen in Section~\ref{sec:requirements}, parallel kernel executions on different PUs can lead to head-of-line blocking (HoL-blocking) and uncertain tail latency for DMA to sNIC/host memory and egress data transfers. For example, kernels can pipeline large storage reads by overlapping asynchronous DMA reads of packet-sized payloads with egress packet sending. OSMOSIS mitigates this by fairly arbitrating IO paths, breaking sizable DMA requests into smaller transactions, and scheduling them with a near-perfect fairness-weighted round-robin (WRR) policy. FMQs supply DMA and egress engines with tenant IO priorities for initiated IO requests. This ensures each tenant obtains a priority-based fair bandwidth chunk.

\subsection{Flexible software control plane}\label{sec:osmosis_api}

OSMOSIS offers a host OS API for sNIC packet processing management, encompassing ECTX creation and offloading specific flow handling to the sNIC. Tenant-initiated offloading involves the creation of a flow ECTX. ECTX facilitates tenant control using the following components.

\subpar{0.5}{SLO policy:}The SLO policy sets compute, DMA, and egress priorities, kernel cycle budget, packet buffer size, and on-sNIC memory. OSMOSIS offers transparent SLO management via SLO knobs indicated in Table~\ref{tab:osmosis_principles}. By default, all tenants' FMQs share equal priority. To achieve perfect fairness in such a scenario, all flows should get the same portion of PUs and IO bandwidth at any time. Increasing the priority of the ECTX leads to \textit{proportionally} more resources (PUs, bandwidth) allocated to the ECTX. A per-kernel cycle limit is adjustable for total or individual kernel execution times and curbs excessive PU usage. Cycle-limit also prevents users from writing ill-behaved code (e.g., infinite \verb|while(true)| loop). We assess SLO's impact on resource fairness in Section~\ref{sec:evaluation}.

\subpar{0.5}{Kernel binary:}kernel binary cross-compiled by the tenant is loaded into sNIC memory by the control plane and is later executed on the flow packets. The kernel binary can compute and schedule DMA and egress requests according to the tenant requirements.

\subpar{0.5}{A virtualized sNIC device:}A virtualized device is allocated for the tenant, e.g., SR-IOV Virtual Function (VF). OSMOSIS associates an IP address with the VF and uses it later for matching, i.e., the VF is 1:1 associated with a single FMQ. Similarly, FMQ-based management can be exposed through any other sNIC virtualization interface, e.g., ~\cite{kumar2019picnic}, ~\cite{justitia}.

\subpar{0.5}{A matching rule:}The matching rule matches packets from the sNIC inbound stream to the ECTX and manages their processing within the same FMQ. A matching rule allows the tenants to open multiple ports on the same virtualized device. The matching engine can match packets based on their UDP/TCP header contents. For example, it can match the IP address and the destination port of the application.

\subpar{0.5}{sNIC memory segments:}The sNIC memory segments are allocated statically to each kernel depending on the requested memory size. The kernels can store the application state in sNIC local memory, e.g., KVS-cache or packet filter table. An error is returned if the tenant uses too much memory or the kernel binary is larger than the SLO policy limits.

\subpar{0.5}{Host memory pages:}The ECTX specifies which host pages can be accessed from the specific kernel via DMA. The DMA engine on the sNIC interfaces the host memory with an IOMMU, translating host virtual addresses to physical addresses. The IOMMU also checks whether the sNIC is accessing an allowed memory region. The control plane initializes the IOMMU with appropriate page tables during execution context creation.

\subpar{0.5}{Event queue (EQ):} An event queue allows the user application to track events like kernel execution errors. When an error occurs (e.g., illegal memory access or exceeding execution time), OSMOSIS informs the host via an event in the kernel's ECTX EQ. A host OSMOSIS API call from the application checks this queue for error messages. EQ can be realized as contiguous sNIC memory mapped to the host virtual address space, akin to RDMA Verbs API EQ~\cite{ibspec}. EQ control path traffic shares the sNIC DMA data path (e.g., PCIe or CXL) with regular kernel execution (e.g., DMA initiated within the kernel) but gets the highest IO priority due to tenants' immediate action needs.

\subsection{Hardware data plane}\label{sec:scheduler}

OSMOSIS provides low management overhead with a minimal hardware footprint. We present two key mechanisms that help us to achieve this goal: a hardware flow abstraction (FMQs) and scheduling algorithms suitable for hardware implementation (WLBVT and DWRR).

\subpar{0.5}{Flow management queues} (FMQs) generalize a packet flow similarly to how a hardware thread generalizes a process. If the tenant needs to offload multiple workloads, each workload kernel (binary) must be associated with its own FMQ. FMQs store matched packet descriptors in a FIFO queue and monitor the flow processing performance. The scheduler then uses these measures to allocate compute resources fairly and enforce per-flow priorities. Processing the FIFO queue triggers kernel executions on sNIC PUs, resembling program instruction execution flow in traditional OS processes.

FMQs also store part of the ECTX state, such as the matching rule, pointers to the kernel binary, and the SLO policy definition. The host-side control plane manages and initializes FMQs that appear as MMIO registers in SR-IOV VF address space. FMQs are highly extensible. For example, the OSMOSIS priority model is compatible with datacenter Ethernet~\cite{pfc}. In case of congestion on the FMQ FIFO queue, the packets can be marked with the appropriate Ethernet ECN congestion flag or can supply the per-FMQ telemetry information~\cite{ibspec,agrawal2020intel,floyd1993random,zhu2015dcqcn,alizadeh2010dctcp,li2019hpcc}.

\subpar{0.5}{FMQ Scheduler}allocates PUs across flows with different compute, DMA, and egress costs-per-packet that are not known \textit{a priori}. Thus, to achieve fair compute utilization, the FMQ arbitration policy needs to be \textit{invariant to the cost-per-byte of the packet} (see Figure~\ref{fig:hpu_unfair}). OSMOSIS implements a hardware scheduler as simple and scalable as the deficit-weighted round-robin (DWRR) but with a minimal additional area footprint (see Section~\ref{sec:implementation}).

OSMOSIS introduces a greedy \textit{Weight Limited Borrowed Virtual Time} (WLBVT) policy, a hybrid of the Weighted Fair Queuing (WFQ) model of FMQ weights and Borrowed Virtual Time (BVT) scheduler. We adopt the BVT algorithm to suit sNIC hardware implementation constraints~\cite{duda1999borrowed, kaffes2019shinjuku} and present our scheduler in pseudo-code Listing~\ref{lst:bvt}. Intuitively, our scheduler aims to allocate each tenant the same amount of PU processing time normalized by priority while ensuring that each tenant is served fairly during PU contention.

An FMQ is in an active state if it contains packet descriptors in the FIFO queue or if its packets are currently being processed on any PU. Flow throughput is updated (\verb|update_tput|) at each sNIC clock cycle only if the corresponding FMQ is active. The scheduler (\verb|get_fmq_idx|) returns the index of the non-empty FMQ that fits the upper limit of weighted PU occupation (\verb|pu_limit| called in line 21) and has the lowest current throughput normalized by FMQ priority (lines 22, 23). 

\begin{lstlisting}[language=Python, label={lst:bvt}, caption=WLBVT FMQ scheduler procedural pseudocode.]
def pu_limit(ActiveFMQs, fmq):
  prio_sum = 0
  for fmq in FMQs:
    if not fmq.empty:
      prio_sum += fmq.prio
  return ceil(len(FMQs) * fmq.prio / prio_sum)

def update_tput(FMQs): #called at each clock cycle
  for fmq in FMQs:
    fmq.total_pu_occup += fmq.cur_pu_occup
    if not fmq.empty or fmq.cur_pu_occup > 0: 
      fmq.bvt += 1 # update only in active state
    fmq.tput = fmq.total_pu_occup / fmq.bvt

def get_fmq_idx(): #called once PU core is free
  min_tput = MAX_INT
  for fmq in ActiveFMQs:
    if fmq.pu_occup < pu_limit(activeFMQs, fmq):
      if fmq.tput / fmp.prio < min_tput:
        min_tput = fmq.tput / fmq.prio
        fmq_idx = fmq.idx
  return fmq_idx
\end{lstlisting}

The weighted PU occupation's upper limit guarantees fair QoS for tenants based on their priority. \verb|pu_limit| is calculated with a \textit{ceil} function to ensure fairness in case of more active FMQs than PUs or non-integer division. The lowest priority normalized throughput equalizes access to oversubscribed PUs over time, favoring users utilizing fewer resources. Our approach can also accommodate total virtual time per tenant (i.e., line 21), which could be useful for billing purposes, thus expanding policy flexibility.

\subpar{0.5}{Kernel execution}is a \textit{short-lived} event as each execution only processes one packet. In OSMOSIS, we run kernels to completion\cite{belay2014ix,prekas2017zygos}. We avoid context-switching for several reasons. As shown in Table~\ref{tab:ctx_lat}, context switching can introduce significant overhead. It also increases the complexity of the hardware data path and requires an additional state per each active kernel.

\subsection{Discussion}
\subpar{0.5}{Run-to-completion model:} If a kernel exceeds a set time limit (e.g., per-FMQ watchdog timer), it's terminated with a hardware interrupt, and the host application receives notification via the corresponding EQ. In this light, we believe that run-to-completion semantics underpins the sNIC programming model that, together with OSMOSIS fair priority adjusted schedulers, ensures predictable packet processing tail latency and also excludes compute-intensive tasks better suited for GPUs or FPGAs. Assuming that the datacenter operator doesn't know the details of the tenant's code to be executed on the PU, the run-to-completion model also ensures the prevention of the execution of ill-behaved code (e.g., a kernel that contains infinite loops).

\subpar{0.5}{Virtual memory:}In principle, OSMOSIS could give each flow the illusion of infinite virtual memory using paging. However, this has two problems. First, translating the page's virtual address to a physical address is a combinational logic operation that will increase the latency of each memory access by at least one cycle. In PsPIN, the on-path backend of OSMOSIS that we discuss in Section~\ref{sec:implementation}, accesses to the L1 scratchpad memory require only $1$ cycle. Second, when using demand paging, swapping in and out memory pages (e.g., between the NIC and the host) also introduces some latency. Because kernels are not context-switched, the PU would actively wait for the page to be swapped in, thereby wasting a large part of the cycle budget.

\subpar{0.5}{Congestion management:} We assume that OSMOSIS is deployed within the lossless network (e.g., InfiniBand, RoCEv2), and FMQs never drop packets. By design, OSMOSIS is compatible with conventional congestion signaling (e.g., ECN) and flow control mechanisms (e.g., Ethernet DCB) supported by existing lossless fabrics. It can also be deployed with DCQCN~\cite{zhu2015dcqcn} and DCTCP~\cite{alizadeh2010dctcp}. From the transport protocol perspective, the packet queueing delay within the FMQs and the corresponding execution of the packet kernel is just another source of latency. For example, the FMQ abstraction deployed with Ethernet can support RED/ECN marking~\cite{ibspec, floyd1993random}. Another mechanism that FMQs can easily support is supplying the P4 INT-MD telemetry information~\cite{agrawal2020intel} to enable the HPCC protocol \cite{li2019hpcc}.

\subpar{0.5}{Encrypted traffic and compute accelerators:}The sNIC handles data movement and may also require accessing the packet contents. Hence, it should be able to decrypt packets (e.g., QUIC~\cite{yang2020making}). sNICs can support either per-PU cryptographic accelerators (e.g., Intel AES-NI~\cite{hofemeier2012introduction}) or a shared accelerator for efficiency (e.g., like in Marvell LiquidIO~\cite{liquidio}) exposed via ISA extensions. In the latter case, the accelerator arbitration resembles PUs, making WLBVT scheduling suitable for compute resource management.

\subpar{0.5}{IO security:}Host memory is protected against unauthorized DMA transfers using an IOMMU setup by OSMOSIS when the host creates the flow context. Similarly, local sNIC memory accesses need to be protected. This can be achieved, for example, by a \textit{Physical Memory Protection} unit (PMP)~\cite{Waterman:EECS-2016-129} as shown in Section~\ref{sec:implementation_PsPIN}.

\section{Implementation}\label{sec:implementation}
We implement OSMOSIS atop PsPIN~\cite{hoefler2017spin,di2021risc}, an open-source on-path sNIC. We adopt PsPIN as a backend for performance-critical operations within OSMOSIS by extending its host-side API to support multiple ECTXs and specify tenant SLOs using $335$ lines of code (LOCs) in C. 

We integrated functional blocks of OSMOSIS (i.e., matching engine, WLBVT scheduler, and DMA request fragmentation) written in $1216$ LOCs of C++ with cycle-accurate simulation PsPIN SystemVerilog backend. In addition, we also implemented these components as synthesizable SystemVerilog IP blocks for hardware cost estimations. These open-source blocks can serve as a future prototype for ASIC or FPGA-based implementation of OSMOSIS.

\subsection{Implementing OSMOSIS on top of PsPIN}\label{sec:implementation_PsPIN}

\subpar{0.5}{Packet processing units:}OSMOSIS PsPIN architecture is based on scalable silicon-proven RISC-V PULP SoC~\cite{di2021risc, kurth2021open,rossi2015pulp}. The PUs are RI5CY 32-bit cores organized in clusters. Each PsPIN cluster contains $8$ PUs clocked at 1GHz and coupled with a $1$ cycle, multi-banked local scratchpad memory (referred to as \textit{L1}). For our experiments, we use the default configuration of the PsPIN PU cluster with 1\,MiB L1 data, and $4$\,KiB L1 instruction caches. Clusters share a global $4$\,MiB L2 packet buffer and a $4$\,MiB L2 kernel buffer, which can be used for local data storage.

\subpar{0.5}{Portable programming API:}OSMOSIS utilizes PsPIN infrastructure to offload the packet processing to the PUs. The user writes a C kernel cross-compiled on the host for the RISC-V ISA architecture. The kernels are then loaded and executed on the flow packets according to the sPIN API~\cite{hoefler2017spin}. PsPIN has a low-latency kernel invocation mechanism ($\leq10$ cycles), i.e., each PU executes a loop polling for a function pointer with the address of the kernel and flow context.

\subpar{0.5}{Kernel IO:}The PsPIN API enables blocking and non-blocking IO calls within kernel code. The PsPIN cluster scratchpad memory is interconnected with the sNIC L2 kernel buffer, host DMA engine buffer, and sNIC egress engine buffer through the 512-bit AXI DMA link. This setup enables \verb|read| and \verb|write| transfers between these buffers, with PUs accessing other cluster memories and shared L2 kernel memory in $10$ to $30$ cycles. This design also transparently supports sNIC egress packet \verb|send|: a DMA \verb|write| from kernel scratchpad memory to the NIC egress engine buffer. PU core L1 scratchpad interfaces an Ethernet egress pipeline over the AXI protocol. PsPIN IO-calls configure a DMA command with addresses, length, and a completion handle pointer. The cluster command FIFO queues outstanding IO commands, and a WRR policy arbitrates per-cluster queues for DMA engine access.

\subpar{0.5}{Memory management:} PsPIN allows specifying the size of the contiguous L1 and L2 memory regions allocatable to tenants' kernels and supports memory isolation using the Physical Memory Protection (PMP) unit. When the kernel accesses L1 and L2 memories, the virtual memory addresses are translated to physical addresses with relocation registers. The PMP then checks that the addresses are within the valid segment range. Like the relocation registers, the PMP unit does not increase the memory access latency~\cite{di2021risc}.

\subsection{OSMOSIS Schedulers}
\subpar{0.5}{FMQ scheduling implementation:}FMQ encompasses a FIFO queue, ECTX (detailed in Section~\ref{sec:osmosis}), and scheduling state. The FIFO queue holds packet descriptors, each containing a 32-bit pointer to the packet. The scheduling state includes a BVT counter tracking tenant resource use and a priority. We implemented the counter as a 64-bit register to avoid overflow\footnote{The 64-bit counter overflow with updates done every cycle at 1 GHz will happen in $2^{64} \div 10^{-9}$s/op $\div 60$s $\div60$m $\div 24$h $\div 365.25$d $\approx584$yrs.}. A 16-bit register stores the FMQ priority. Our SystemVerilog WLBVT implementation with 128 FMQs synthesizes at 1 GHz, making a scheduling decision in five cycles. Most latency stems from the weight-limiting requiring integer division, which is challenging for fast hardware implementation. We hide this latency using pipelining, overlapping FMQ arbitration with packet DMA from the L2 packet buffer to the cluster scratchpad (at least 13 cycles for a 64-byte packet).

\subpar{0.5}{Enhanced DMA engine:} To prevent HoL-blocking, OSMOSIS applies transfer fragmentation on both the host-interfacing DMA engine and the egress engine. We implement two modes of fragmentation: a \textit{software} fragmentation implemented within the kernel call for a DMA transfer and a \textit{hardware} fragmentation within the DMA engine. The software approach wraps \verb|pspin_dma_read/write| and \verb|pspin_send_packet| with a function, dividing larger requests into smaller chunks. We issue multiple non-blocking DMA requests of smaller sizes while internally maintaining the state for each transfer. While this optimization mitigates HoL-blocking (as shown in Section~\ref{sec:evaluation}), it also hinders the throughput of large DMA requests. To minimize this, we expand the functional model of AXI to enable hardware DMA fragmentation offloading. This involves managing the state for multiple outstanding AXI write requests and arbitrating them with the WRR scheduler.

\subsection{Integration with other on-path SmartNICs}\label{sec:bf3dpa}

OSMOSIS could be applied to the on-path sNIC designs besides PsPIN. For example, the data path accelerator (DPA) introduced in Bluefield 3 DPU~\cite{bf3-flexio, bf3} could be extended with OSMOSIS to enable kernel execution QoS. DPA invokes user-defined kernels upon completion of RDMA operations. Thus, FMQ abstraction could be 1:1 mapped to DPA-managed RDMA Completion Queues (CQs) that are arbitrated according to OSMOSIS WLBVT policy. Further, IO operations initiated from DPA cores during kernel execution, i.e., RDMA Work Requests (WRs), could be assigned with a desired Service Level (SL) mapped to the underlying RDMA Virtual Lane (VL), i.e., SL2VL mapping mechanism~\cite{ibspec}.

\section{Evaluation}\label{sec:evaluation}

We study how OSMOSIS allocates sNIC resources under different traffic conditions and workload requirements. We investigate the following research questions:
\begin{enumerate}
    \itemsep0em
    \item How does the area of OSMOSIS-enabled sNIC chip scale up with the ingress link rates and the number of tenants?
    \item What are the overheads of OSMOSIS compared to the reference PsPIN implementation?
    \item What is the maximum load that OSMOSIS can sustain?
    \item How fair are OSMOSIS resource allocations?
\end{enumerate}

\subsection{Hardware Scaling}

\begin{figure}[]
    \centering
    \includegraphics[width=1.0\columnwidth]{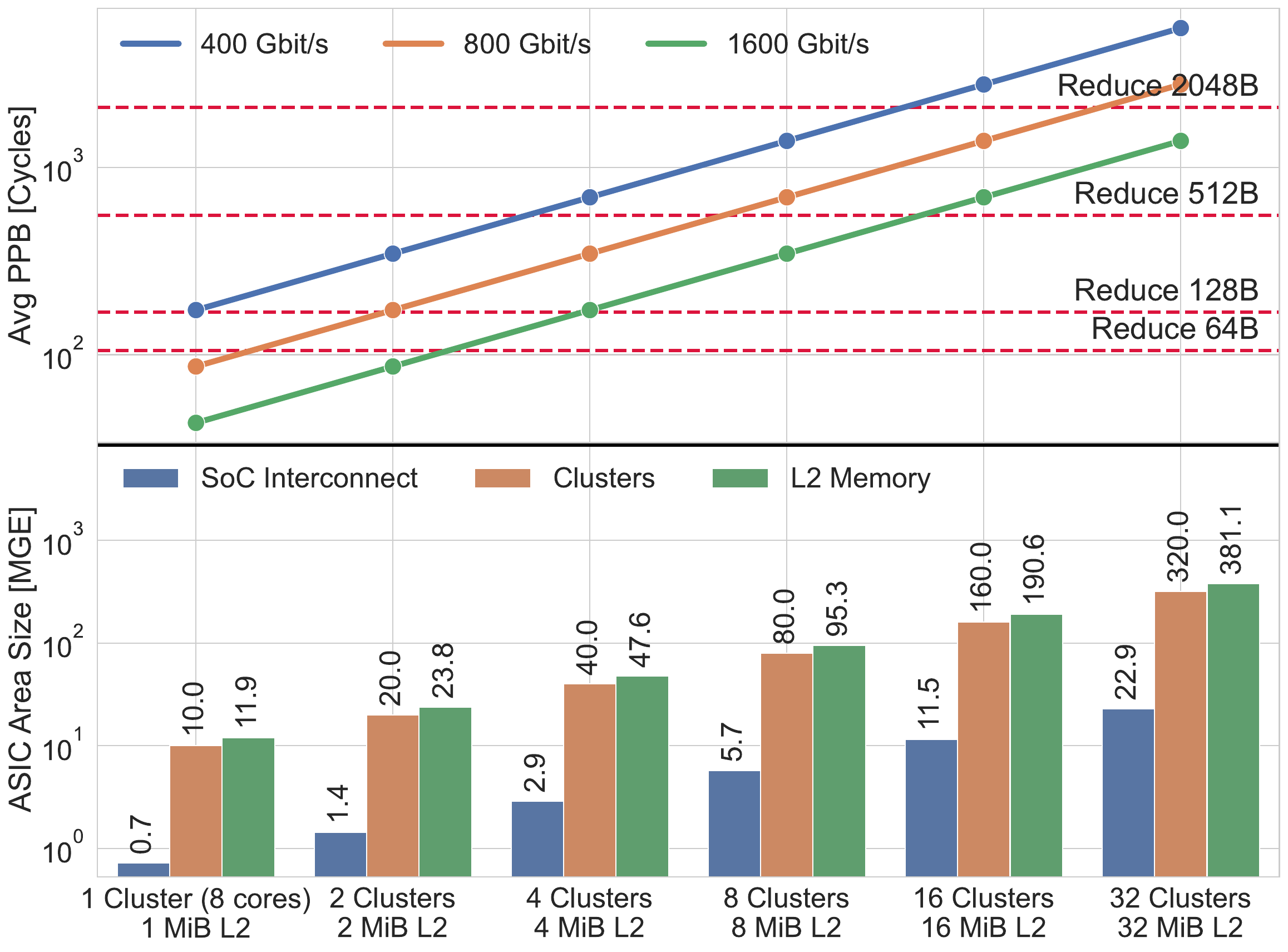}
    \caption{The cost model of sNIC SoC area synthesized in 22nm GF process, compared to the theoretical per packet budget (averaged for different packet sizes at 64 -- 4096 B interval) achieved with 400/800/1600 Gbit/s ingress link rates.}
    \label{fig:cluster_silicon_scaling}
\end{figure}

We synthesized OSMOSIS and PsPIN SystemVerilog IP blocks at 1GHz in GlobalFoundries 22nm node process to estimate hardware area costs using Synopsys Design Compiler NXT in topographic mode.

\subpar{0.5}{sNIC area scaling with compute capacity:} PsPIN clusters utilize a hierarchical SoC-interconnect similar to Manticore scale-out study~\cite{zaruba2020manticore}. We group four clusters into a \textit{quadrant} sharing a local interconnect. Each quadrant connects to L2 memory, allowing all cores to access the shared packet buffer. Synthesis studies~\cite{kurth2021open, di2021risc} indicate negligible area increases and timing overheads when adding ports to L2. In Figure~\ref{fig:cluster_silicon_scaling}, PsPIN demonstrates linear compute capacity scaling relative to the core area. For instance, 4 PU clusters offer adequate per-packet budget (PPB) (Section~\ref{sec:requirements}) to sustain compute-bound Reduce workload with up to 512-byte packets.

\begin{figure}[]
    \centering
    \includegraphics[width=\columnwidth]{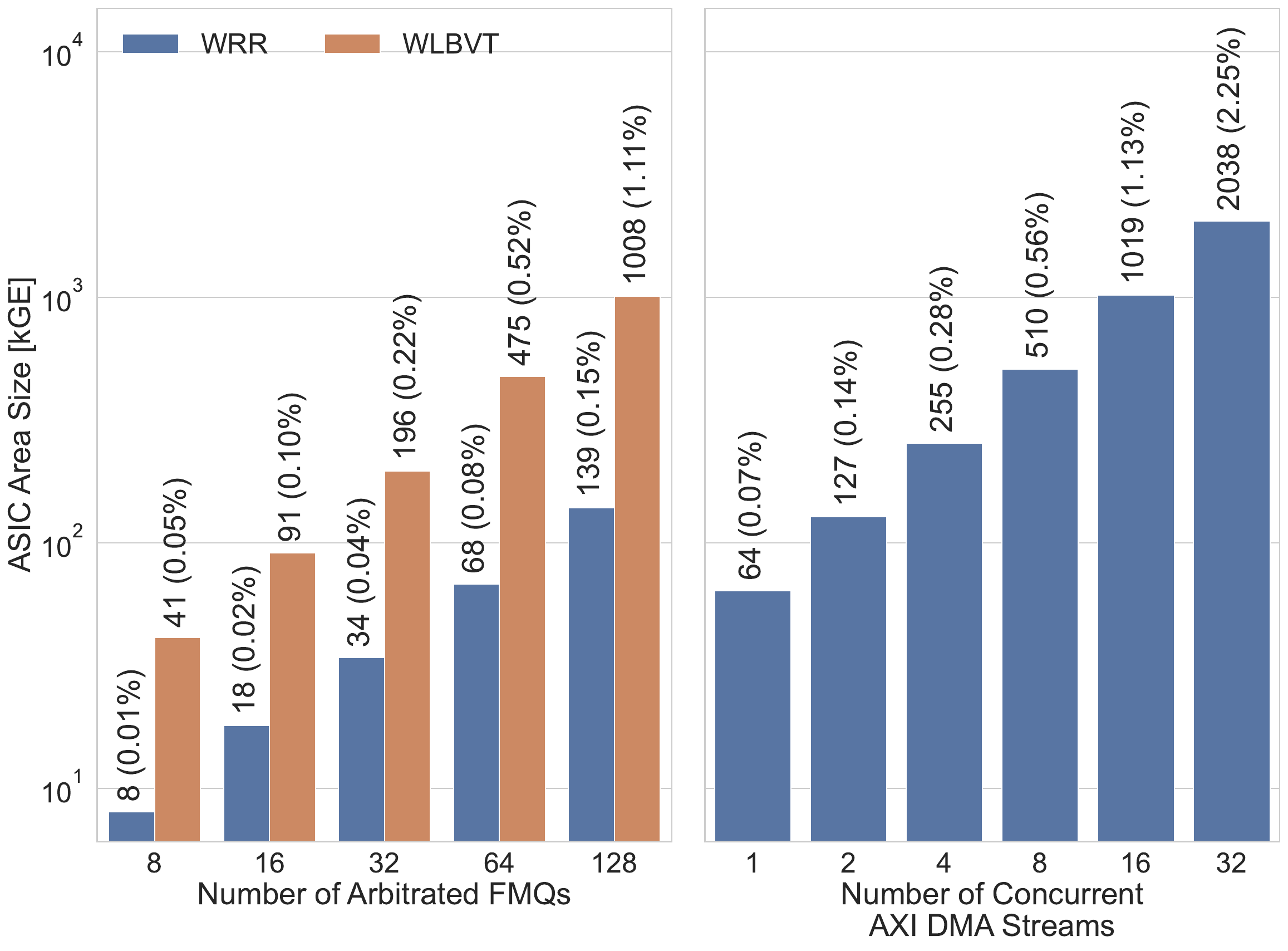}
    \caption{WLBVT and WRR exhibit linear area scaling in the GF 22nm process. Bar captions indicate gate count and relative area compared to 4 PU clusters with 4 MiB L2.}
    \label{fig:scheduler_silicon_scaling}
\end{figure}

\subpar{0.5}{OSMOSIS Schedulers Scaling:}Figure~\ref{fig:scheduler_silicon_scaling} shows the hardware area consumption of OSMOSIS schedulers. We observe a linear scaling of the FMQ and DMA engine schedulers with the number of inputs. Assigned with a custom packet matching rule, one FMQ scheduler input can serve millions of requests, such as independent IO reads/writes (see Figure~\ref{fig:raw_apps}). Compared to RR, WLBVT needs $7\times$ more gates, yet with 128 FMQs, WLBVT area consumption takes only $1\%$ of PsPIN cluster and L2 memory area. With a reasonable hardware footprint, OSMOSIS enables the hardware scheduling of up to 128 tenants subscribed to the same SmartNIC.

\subsection{Experimental Methodology}

We evaluate OSMOSIS runtime performance using cycle-accurate Verilator v4.228 SystemVerilog simulator~\cite{snyder2013verilator}. Our experimental testbed features two setups: a \textit{Reference (baseline) PsPIN implementation}, i.e., a conventional on-path sNIC without multi-tenant OS, and a \textit{PsPIN implementation enhanced with OSMOSIS management}.

Both setups feature 4 PsPIN clusters of $8$ $1$GHz cores, achieving $400$ Gbit/s ingress/egress bandwidth. L2 and host memories can be accessed through a $512$ Gbit/s AXI link. We used randomly pre-generated packet traces that fully saturate ingress link bandwidth. Packet arrival sequences follow a uniform distribution, and packet sizes are sampled from a lognormal distribution~\cite{clusterdata:dcburst,clusterdata:fbdataset, clusterdata:private}. For fairness measurements, we use Jain's fairness metric~\cite{hossfeld2016definition}. It scales between 1 and 1 divided by the number of tenants: a metric of $y$ implies $y$\% fair treatment, leaving $(100-y)$\% starved. Fair treatment ensures equal priority-adjusted resource access for each tenant.

A RR scheduler is available in the reference PsPIN implementation, thus we consider it as a baseline. To our knowledge, production on-path designs (e.g., BlueField-3 DPA) use static compute management. We consider a dynamic scheduler over static allocation for the baseline since work conservancy is an essential requirement for datacenter energy efficiency. We discuss how OSMOSIS differs from existing NIC management solutions in Section~\ref{sec:comparison}.

\subsection{Synthetic Benchmarks}

\label{sec:synthetic_benchmarks}
We evaluate OSMOSIS on synthetic benchmarks to assess its overheads in a low-complexity environment.

\begin{figure}[]
    \centering
    \includegraphics[width=1.0\columnwidth]{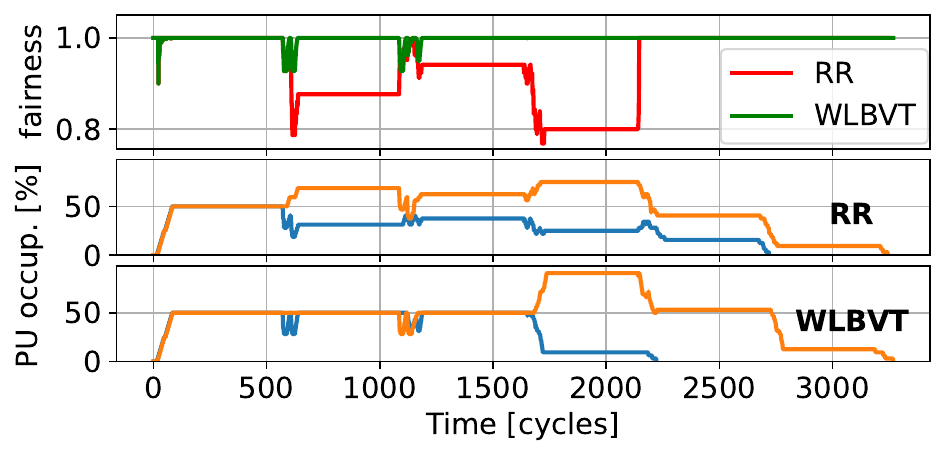}
    \caption{The fairness of WLBVT and RR with two tenants of different compute cost per byte. The orange line in the PU occupation subplots represents the \textit{Congestor} tenant, whose workload consumes $2\times$ more cycles per packet than the \textit{Victim} tenant depicted by the blue line.}
    \label{fig:fair_pu_utilization}
\end{figure}

\subpar{0.5}{\negcircnumreq{R1}\negcircnumreq{R5} Fair HPU allocation:} We run two applications, one with a larger \textit{compute cost per byte}, the \textit{Congestor}, and the other with a smaller one, the \textit{Victim}. Both spin in a \verb|for| loop to simulate a compute-bound task. Figure~\ref{fig:fair_pu_utilization} shows how RR over-allocates PUs to the \textit{Congestor}, leading to lower fairness, as shown by Jain's metric. WLBVT consistently splits all the resources equally between tenants. When the \textit{Victim} has no outstanding packets, WLBVT allows the \textit{Congestor} to overtake more PUs. WLBVT enables fair compute resource allocation within OSMOSIS and does not cause slowdowns within the benchmarks.

\begin{figure}[]
    \centering
    \includegraphics[width=1.0\columnwidth]{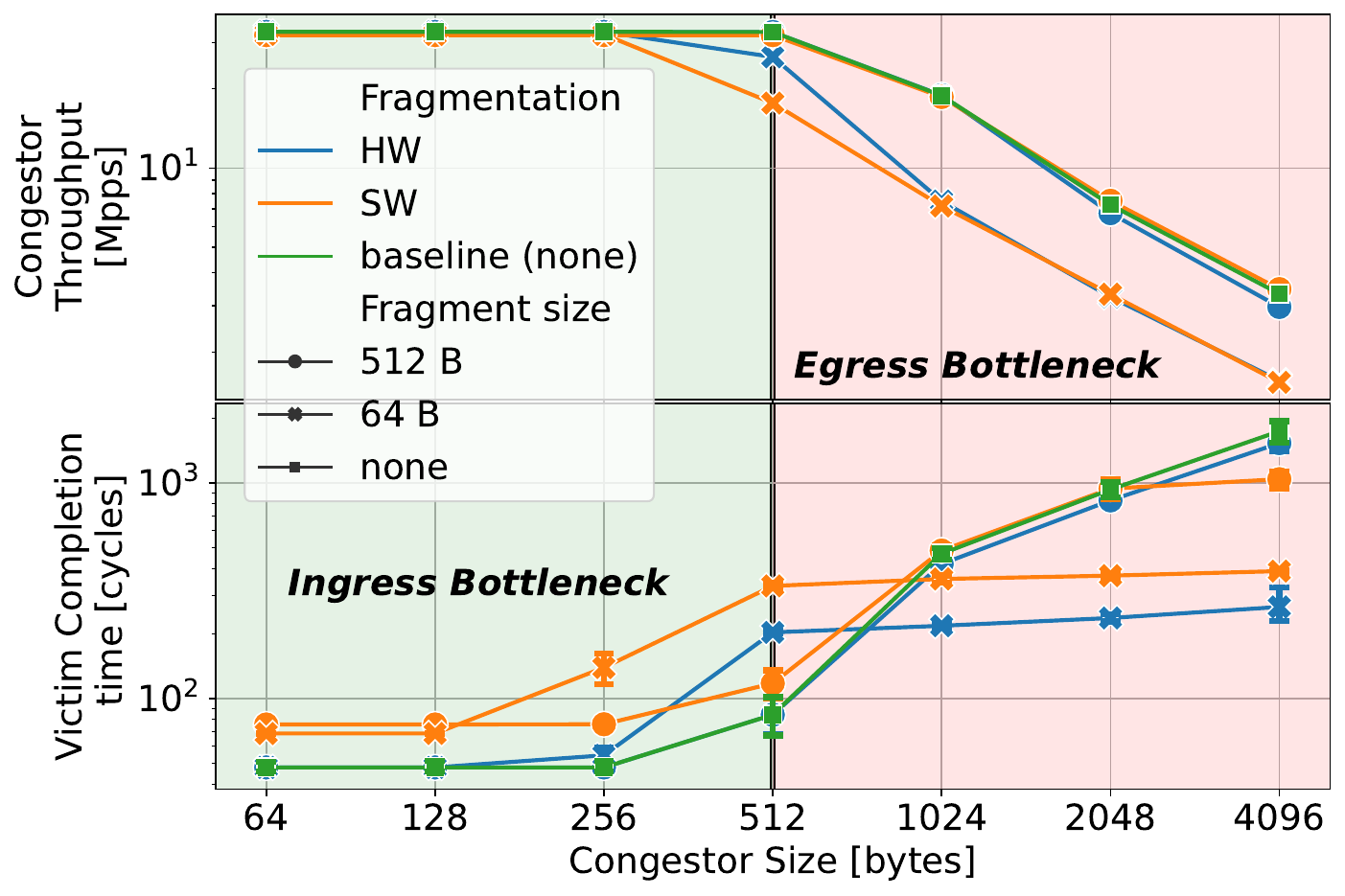}
    \caption{The impact on the \textit{Congestor} throughput and the \textit{Victim} kernel completion time as a function of the \textit{Congestor} size and various fragment sizes. State transition between ingress and egress bottleneck depicts where the line rate of the egress path became saturated.}
    \label{fig:dma_hol_fixed}
\end{figure}

\subpar{0.5}{\negcircnumreq{R2}\negcircnumreq{R5} Resolving HoL-blocking:}We evaluate the scaling of throughput of the \textit{Congestor} and the kernel completion time of the \textit{Victim} while conducting only Egress transfers that involve AXI writes. Figure~\ref{fig:dma_hol_fixed} presents how OSMOSIS resolves HoL-blocking. Depending on the fragmentation method, the \textit{Victim}'s kernel completion time can be reduced by an order of magnitude while preserving a relative slowdown of only around 2$\times$. The throughput reduction stems from control traffic overhead related to fragmentation, i.e., splitting one large transfer into smaller $N$ transfers introduces $N$ additional protocol handshakes between sender and receiver. When accessing local sNIC memories (i.e., remote scratchpads and L2), it can be mitigated through a custom SystemVerilog implementation of the PsPIN AXI protocol, allowing for parallel transfer states as proposed in other works~\cite{benz2023high, restuccia2020axi, jiang2022axi}. Addressing this issue for host-side traffic that crosses AXI bus boundaries would require a fine-grained QoS protocol for PCIe and CXL interconnects~\cite{agarwal2022understanding}. 

We also observed two bottlenecks: ingress and egress. In the ingress bottleneck, the incoming link bandwidth is the limit, while in the egress one, the AXI bus congestion causes slowdowns. While the overheads come from the interconnect, OSMOSIS scheduling does not introduce overheads, as evident for low \textit{Congestor} sizes.

\subsection{Datacenter Workloads}\label{sec:datacenter_workloads}
Additionally, we evaluate a set of real datacenter workloads supplied with the PsPIN benchmarking package~\cite{di2021risc}. We study the \textit{Aggregation}~\cite{ordonez2011horizontal}, \textit{Reduction}~\cite{ben2019demystifying} and \textit{Histogram}~\cite{barthels2017distributed} benchmarks as examples of compute-bound workloads with incrementally increasing inter-kernel memory synchronization requirements, i.e., from local on-PU computation with one atomic operation in \textit{Aggregation}, to random memory accesses, each with an atomic summation in \textit{Histogram}. 

We also evaluate an IO-bound benchmarking set. Our goal is to exercise NIC DMA read/write data paths towards the host memory, the pattern typical for data path offloading of storage RPCs and TCP segment delivery~\cite{min2021gimbal,pismenny2021autonomous, moon2020acceltcp, shashidhara2022flextoe}. In \textit{IO read/write} workloads, a target memory location is stored directly in the packet application header. The multiple clients make concurrent IO requests to the same storage node, and we serialize all requests through $1$ FMQ that serves either read or write requests. 

In the \textit{Filtering} benchmark, to lookup the destination DMA memory address (e.g., KVS-cache location or packet forwarding table context address), the kernel needs to compute the hash of the L7-header used as a lookup table index stored in sNIC LLC.

\begin{figure}[]
    \centering
    \includegraphics[width=1.0\columnwidth]{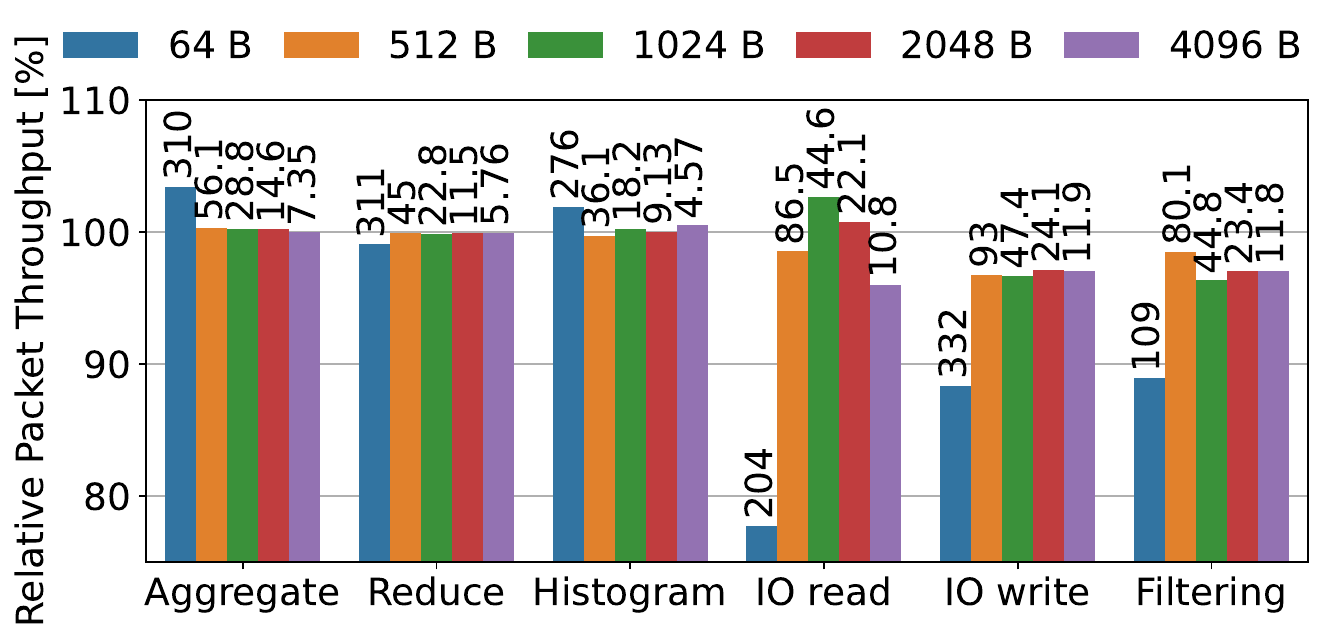}
    \caption{The relative packet throughput of common datacenter workloads run in a standalone mode as a function of packet size with their raw performance in million packets per second (Mpps) at the top of the bars. Up to a $3\%$ throughput increase with OSMOSIS compared to the PsPIN baseline stems from a kernel completion time variability introduced by the compute/IO schedulers.}
    \label{fig:raw_apps}
\end{figure}

\subpar{0.5}{Management overheads:}To assess the influence of OSMOSIS management on applications' performance, we start by running them in isolation. Figure~\ref{fig:raw_apps} displays how OSMOSIS does not introduce considerable overheads for compute-bound workloads. These oscillate within $\pm 3$\% of the baseline PsPIN implementation and reach the maximum of 310Mpps for the \textit{Aggregation} workloads. For IO-bound workloads, OSMOSIS introduces overheads stemming from the fragmentation, which have been discussed in Section~\ref{sec:synthetic_benchmarks}. This can be resolved by extending the AXI bus protocol~\cite{restuccia2020axi, jiang2022axi}. While overheads reach from 23\% to 2\% and represent the cost of introducing fair and efficient multi-tenancy, the workloads still achieve 332Mpps in the \textit{IO write} case. 

\subpar{0.5}{Application mixtures:}Evaluating applications in isolation is not representative of real workloads which occur in multi-tenant datacenters for which OSMOSIS was designed and where multiple users contend for resources. We consider two application sets: \textit{compute} and \textit{IO}, each resulting in tenant resource contention.

\begin{figure*}[t!]
    \centering
    \begin{subfigure}[t]{0.5\textwidth}
        \centering
        \includegraphics[width=1.0\columnwidth]{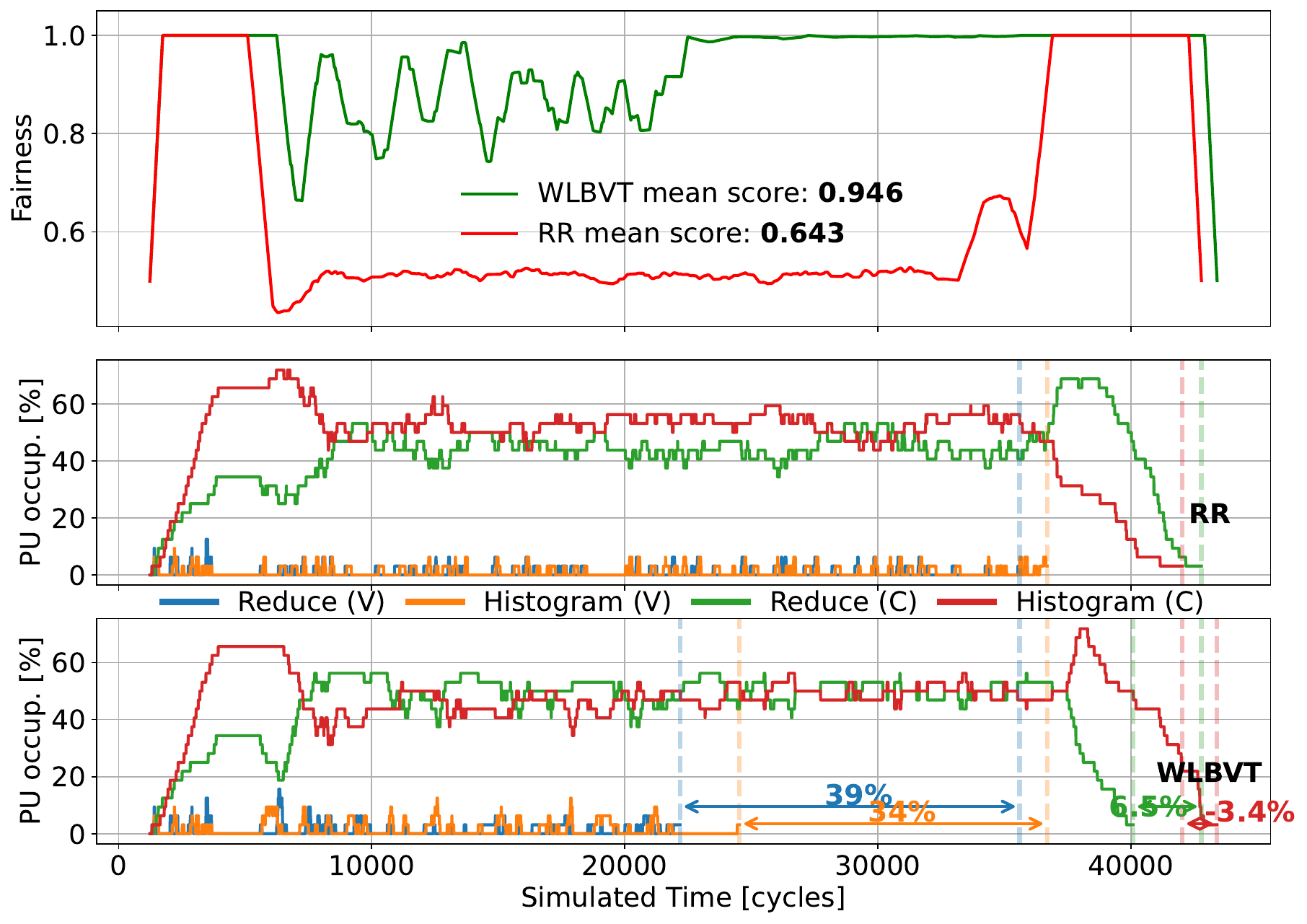}
        \caption{Compute workload set}
        \label{fig:hpu_compute_mixes}
    \end{subfigure}%
    ~ 
    \begin{subfigure}[t]{0.5\textwidth}
        \centering
        \includegraphics[width=1.0\columnwidth]{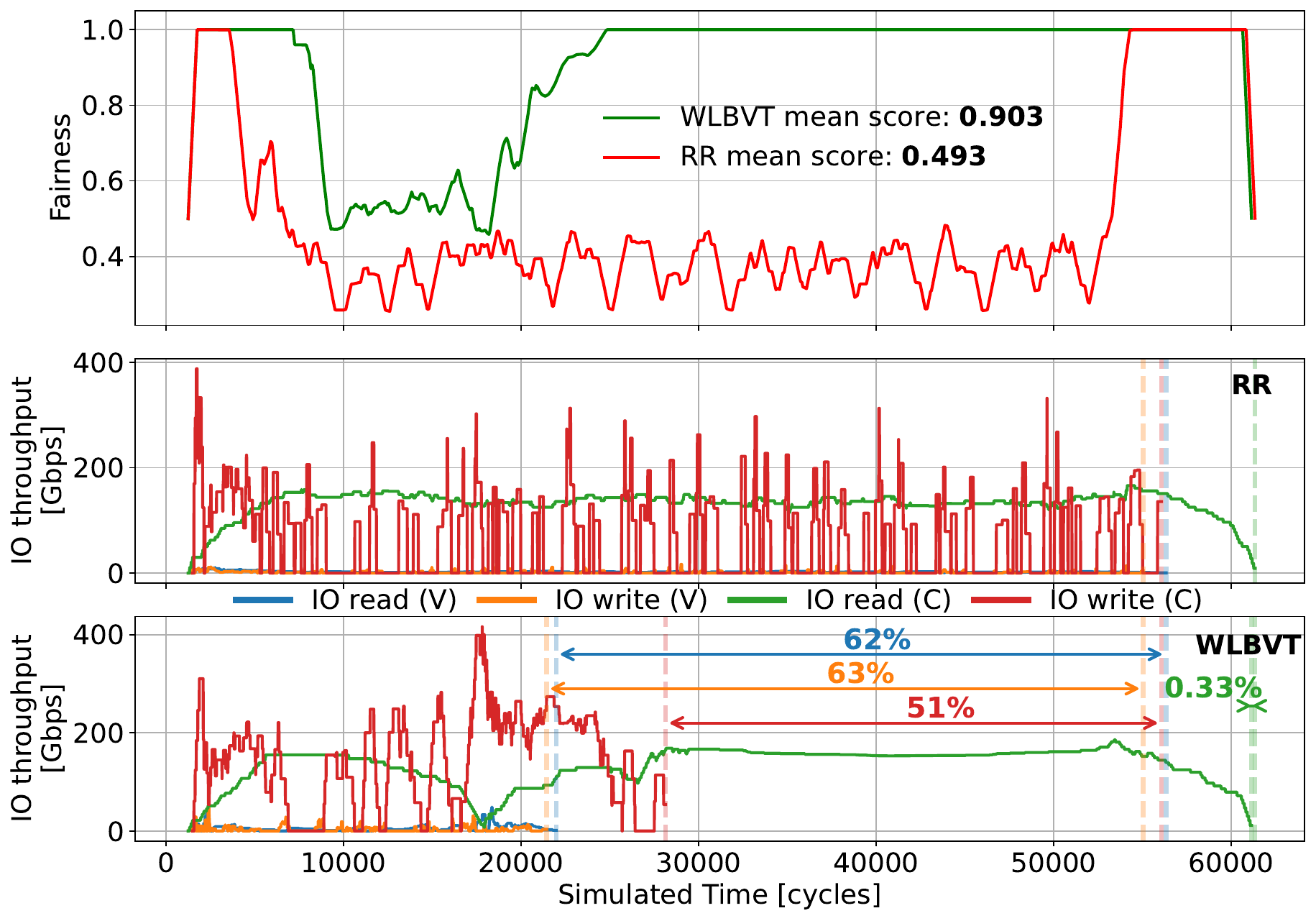}
        \caption{IO workload set}
        \label{fig:hpu_mixes_IO_tput}
    \end{subfigure}
    \caption{The evolution of tenant performance and average fairness against the simulated time. The upper sub-plots show the total Jain's fairness score computed over all flows at once. The percentages indicate the reduction in FCT for each tenant.}
\end{figure*}

The compute-bound set comprises the \textit{Reduce} and \textit{Histogram} workloads. Each is introduced as a \textit{Victim} (64B packets for \textit{Reduce} and 64-128 packets for \textit{Histogram}) and \textit{Congestor} (4KB packets for \textit{Reduce} and $3072$ -- $4096$ byte packets for \textit{Histogram}). As Figure~\ref{fig:hpu_compute_mixes} shows, these workloads saturate the PUs of the sNIC within the first couple thousand cycles and introduce compute congestion. Using OSMOSIS WLBVT scheduling, each tenant obtains an average allocation of 47\% fairer than that of the typical RR implementation as measured using Jain's metric. Such allocations ensure SLO fulfillment and result in 39\% faster \textit{flow completion times} (FCT) because of lower average contention while only sacrificing 3\% of the \textit{Histogram} \textit{Congestor}. OSMOSIS thus achieves a fair and efficient resource allocation.

The multi-tenancy system must efficiently manage \textit{all} resources in coordination~\cite{ghodsi2011dominant, ghodsi2012multi}. We illustrate this scenario in Figure~\ref{fig:hpu_mixes_IO_tput}, where the IO set includes $4$ kernels of varying complexity so that the code executed on the PUs produces various data movement patterns. The set consists of IO \textit{read} and \textit{write} flows, introduced again as both a \textit{Victim} and \textit{Congestor}. While \textit{reads} and \textit{writes} share the NIC ingress, the utilized DMA paths are opposite to each other.

The \textit{write} packets have a variable-length packet size (up to 128B and 4KB for \textit{Victim} and \textit{Congestor}, correspondingly) proportional to the payload size. The payload of the \textit{read} flow has a fixed size and contains $2$ $64$-bit values (read location in memory and its size varied for \textit{Victim} and \textit{Congestor}). While each \textit{read} packet will spend fewer cycles in the NIC ingress, it will induce up to $2\times$ more data movement work compared to \textit{write}, i.e., DMA read from the host memory followed by sending towards egress. This results in seemingly continuous distributions for \textit{read} requests that are processed slower than bursty \textit{writes}.

\begin{figure}[t]
    \centering
    \includegraphics[width=1.0\columnwidth]{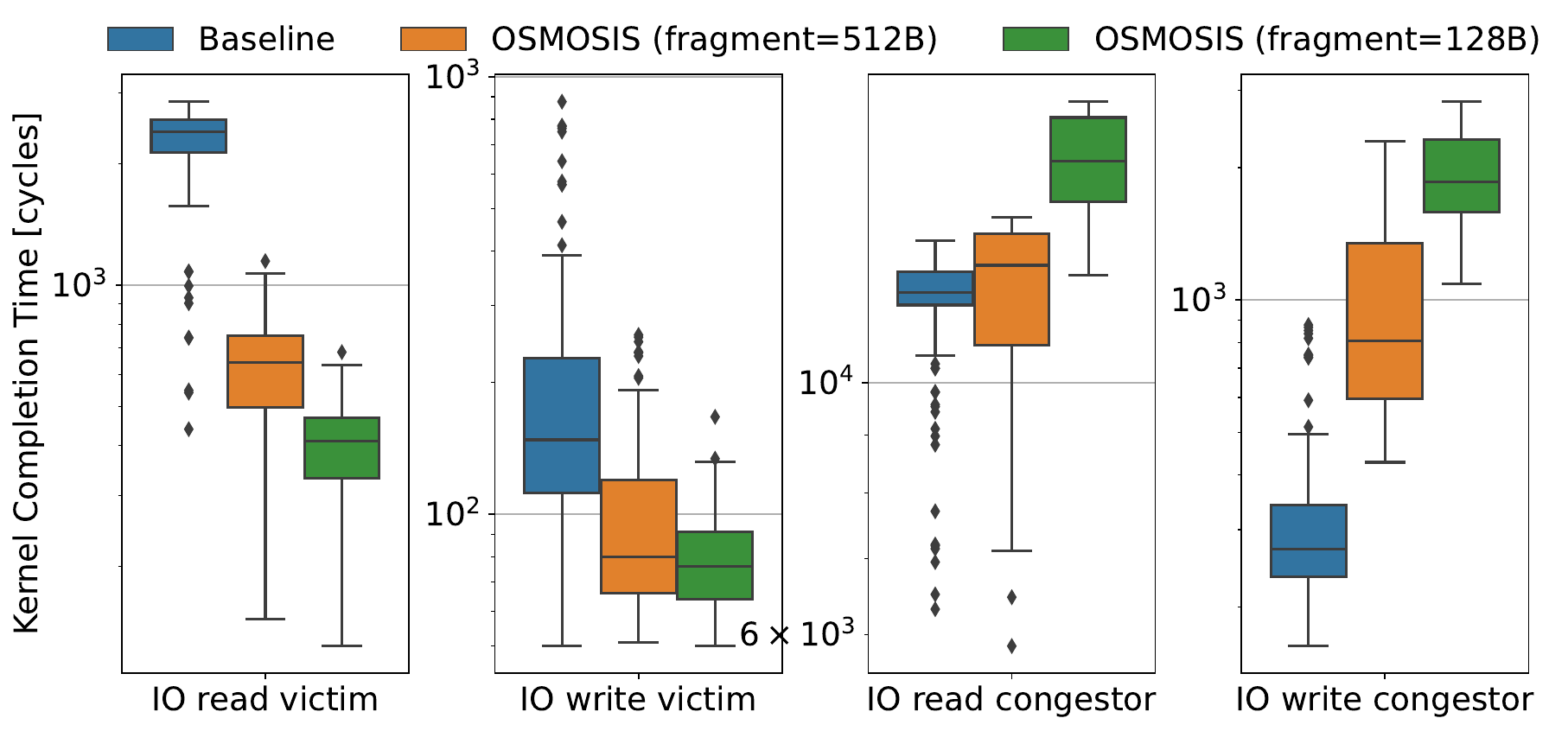}
    \caption{The completion time distribution for IO-bound applications for two fragment sizes.}
    \label{fig:hpu_mixes_IO}
\end{figure}

Figure~\ref{fig:hpu_mixes_IO_tput} shows that, similarly to the compute case, OSMOSIS obtains a consistently fairer allocation than a RR scheduler (up to 83\%) as measured by the average Jain's fairness metric. We notice that the \textit{writes} are processed much faster than the \textit{reads}. OSMOSIS also manages to reduce FCT for all tenants by up to 63\%. Such large improvement comes from addressing the HoL-blocking problem, leading to a more efficient allocation. The \textit{IO read} \textit{Congestor} is initially suppressed to let other tenants fairly finish their workloads and then obtains full exclusive utilization, eliminating contention and allowing it to regain the lost performance. On the other hand, the other tenants are fairly allocated and, as Figure~\ref{fig:hpu_mixes_IO} shows, they do not suffer from HoL-blocking.

Figure~\ref{fig:hpu_mixes_IO} also displays the true cost of the aforementioned gains. While the overall FCT is reduced for all tenants, the single kernel completion time shows a different story. The HoL-blocking is resolved for the \textit{Victim} tenants, for which the kernel completion time is reduced more than fivefold. However, the other \textit{Congestor} tenants display an up to $8\times$ increased median kernel completion time. While OSMOSIS increases the median per packet processing time, it also achieves overall FCT gains for the IO set by allocating the resources fairly, and by parallelizing the packets appropriately. 

\section{Related Work}\label{sec:comparison}

In this section, we summarize recent milestones in NIC resource management research and provide a qualitative comparison of existing solutions with OSMOSIS.

Justitia~\cite{justitia} and PicNIC~\cite{kumar2019picnic} are rNIC virtualization layers lacking on-NIC compute management. They function as software controllers between the NIC and host application, handling RDMA read/write operations atop the RDMA API. Lynx~\cite{tork2020lynx} focuses on sNIC GPU data movement offloading but similarly manages traffic at a \textit{per-message} granularity and lacks detailed analysis of multi-tenancy issues.

Floem~\cite{floem}, FairNIC~\cite{grant2020smartnic}, and iPipe~\cite{liu2019offloading} specifically target on-path sNICs programmability. All three solutions lack flow priorities implementation. FairNIC aims for multi-tenant use cases by statically allocating compute and IO bandwidth to flows. This approach can potentially cause under-utilization or unfairness~\cite{prekas2017zygos,kaffes2019shinjuku,seyedroudbari2023turbo}. iPipe \cite{liu2019offloading} proposes to move the execution of packet processing to the host CPU in case of congested sNIC resources. We design OSMOSIS for scenarios where on-path sNIC \textit{fully} offloads the packet processing, and the host CPU runs a server-local non-networking path on the data processed by sNIC, e.g., computation on the results of in-network reduction or host-local distributed file system management.

Per-flow priority management is present in PANIC~\cite{panic} and Menshen~\cite{menshen}. Both solutions specialize in Reconfigurable Match Tables (RMT) pipeline architectures, e.g., PANIC is tailored for FPGA-based sNICs. The applicability scope of OSMOSIS is different, focusing on programmable on-path designs such as Bluefield-3 DPA and PsPIN, which explore a different type of parallelism. In on-path sNICs the packets of the same flow are processed in parallel with user-defined C kernels. These kernels run on tens to hundreds of energy-efficient cores integrated within one SoC. To efficiently distribute packets across a large core count and sustain the line rate, on-path sNICs are constrained with low-latency hardware packet schedulers lacking reconfigurability.

To our knowledge, OSMOSIS is the first solution that can support fair work-conservative SLO-based traffic management integrated within the on-path sNICs hardware data path.

\section{Conclusions}\label{sec:relatedwork}
Enabling user-level on-NIC processing in modern multi-tenant datacenters brings resource multiplexing and hardware/software co-design challenges. OSMOSIS solves sNIC multi-tenancy by distributing sNIC resources, the egress and DMA bandwidth, and processing units across flows with different priorities, input bandwidth, and computational requirements. To achieve a fair distribution of resources, OSMOSIS relies on sNIC-specific principles, such as work-conservative allocation of compute and IO resources. The evaluation shows that OSMOSIS efficiently redistributes resources, enabling QoS, performance isolation, and prioritization between various mixtures of flows. OSMOSIS improves FCT by up to 60\% and is fairer by up to 83\% than typical schedulers. We believe that OSMOSIS could enable wider adoption of on-path sNICs in cloud datacenters with low overhead.

\vspace{2em}

\section*{Acknowledgments}
This project received funding from EuroHPC-JU under the grant agreements RED-SEA, No. 055776 and DEEP-SEA, No. 95560, the EuroHPC-JU "The European Pilot" project under the grant agreement No. 101034126 as part of the EU Horizon 2020 research and innovation programme, and a donation from Intel.

{\footnotesize \bibliographystyle{acm}
\bibliography{osmosis}}

\end{document}